\documentstyle[emlines,12pt]{article}
\begin{document}

\begin{center}
{\large\bf Nonlinear Boundary Value Problem of Magnetic Insulation}
\end{center}
\begin{center}
{\large\it A.V. Sinitsyn}

Institute System Dynamic and Control Theory of SB RAS,\\
Lermontov Str., 134, 664033 Irkutsk, Russia,\\
 e-mail: avsin@icc.ru
\end{center}
\vspace{0.2cm}

{\bf Abstract.}
On the basis of generalization of upper and lower solution method to the
singular two point boundary value problems, the existence theorem of
solutions for the system, which models a process of magnetic insulation in
plasma is proved.

\vspace{0.3cm}

\vspace{0.5cm}

{\bf1. Introduction.}

Investigation of mathematic models of magnetic insulation has been started by
P.Degond, N.Ben Abdallah and F.Mehats in 1995 year. In 1996 P.Degond has put
to the author of this paper the problem on existence of solutions of limit
system (I) and its generalization to the problem with free boundary. The effect
of magnetic insulation consists in that the electrons emitted from cathode
cannot reach the anode due to the extremely high applied electric and magnetic
field; they are reflected by the magnetic forces back to the cathode. Thus
there is electronic layer outside of which electromagnetic field is equal to
zero [1]. Here two basic regimes are possible: the first, when electrons reach the
anode $-$ "noninsulated" diode and the second one, when electrons rotate back to
the cathode $-$ "insulated" diode. The regime of "noninsulated" diode is
described by the following nonlinear two-point boundary value problem
$$
\frac{d^{2}\varphi}{dx^{2}}=j_{x}\frac{1+\varphi(x)}{\sqrt{(1+\varphi(x))^{2}-
1-a(x)^{2}}}\stackrel{\triangle}{=}F(\varphi,a); \;\;\;\varphi(0)=0,\;\;
\varphi(1)=\varphi_{L},
$$
$$
\eqno {\rm (I)}
$$
$$
\frac{d^{2}a}{dx^{2}}=j_{x}\frac{a(x)}{\sqrt{(1+\varphi(x))^{2}-1-a(x)^{2}}}
\stackrel{\triangle}{=}G(\varphi,a); \;\;\;a(0)=0,\;\;a(1)=a_{L},
$$
where $j_{x}>0$, $x\in [0,1]$; $\varphi$ is the potential of electric field
and the potential of magnetic field is $a$.

Our main goal consists in search of positive solutions of system (I) that is
$\varphi>0$, $a>0$ and their dependences upon parameter $j_{x}$. Here there are
some interesting questions about solvability of this problem, because the
system (I) is singular in zero for $\varphi=0$ and in this connection, we can
not say about properties of monotonicity of right parts on the interval
$\varphi\in [0,\infty)$ and, hence, about Lipschitz condition. The problem (I)
has no a property of quasimonotonicity in cone. Thus a standard upper and lower
solution method, developed for the systems of semilinear elliptic equations
in partially ordered Banach space [3], does not work. In spite of this fact,
we show the existence of lower and upper solutions of problem (I) without
conditions of local Lipschitz continuity and quasimonotonicity using sufficiently
simple techniques. To this purpose, we modify the P.J.McKenna and W.Walter [4]
theorem of existence of lower and upper solutions for arbitrary elliptic
systems
$$
\triangle u+f(x,u)=0 \;\;{\rm в}\;\;\Omega, \;\;\;u=0\;\;{\rm on}\;\;
\partial\Omega,
$$
where $u=(u_{1},\ldots,u_{n})$, $f=(f_{1}, \ldots, f_{n})$ are $n$ $-$ vectors,
$\Omega$ $-$ is an open bounded subset of $R^{M}$ with smooth boundary
$\partial\Omega$, and $f(x,u)$ is uniformly H$\ddot{o}$lder continuous (with
exponent $\alpha$) in $x$ and Lipschitz continuous in $u$.

In section 2 we will prove theorem 2 and propositions 1, 2 on the existence of
semitrivial solutions of problem (I) by upper and lower solution method. The
estimations to the value of electrostatic potential on the anode $\varphi_{L}$
and the current $j_{x}$ are obtained. In section 3 we formulate the principal
theorem 4 on the existence of positive solutions of problem (I) and the
estimation to the value of magnetic field on the anode $a_{L}$ is given.

We note that system (I) was studied in [2] by a shooting method with
$\beta=a'(0)$ and $j_{x}$ as shooting parameters. The strategy is:
given the values of $\beta$ and $j_{x}$, solve (I) with the Cauchy conditions
$\varphi(0)=0$, $a(0)=0$, $\varphi'(0)=0$, $a'(0)=\beta$, and then adjust the
values in order to fulfill the conditions $\varphi(1)=\varphi_{L}$ and
$a(1)=a_{L}$.
\vspace{0.2cm}

{\bf 2. Existence of semitrivial solutions of problem (I).}

Let us introduce the definition of cone in a Banach space $X$.

{\bf Definition 1.} Let $X$ be a Banach space. A nonempty convex closed set
$P\subset X$ is called a cone, if it satisfies the conditions:\\
(i) $x\in P$, $\lambda\ge 0$ implies $\lambda x\in P$;\\
(ii) $x\in P$, $-x\in P$ implies $x={\cal O}$, where ${\cal O}$ denotes
zero element of $X$.\\
$\le$ is the order in $X$ induced by $P$, i.e. $x\le y$ if and only if
$y-x$ is an element of $P$.

We will denote $[x,y]$ the closed order interval between $x$ and $y$, i.e.
$$
[x,y]=\{z\in X: x\le z\le y\}.  \eqno(1)
$$
We will also assume that the cone $P$ is normal in $X$, i.e. order intervals
are norm bounded.

In $X$
$$
X\equiv\{(u,v):\;u,v\in C^{1}(\bar{\Omega}), u=v=0\}
$$
we introduce the norm $|U|_{X}=|u|_{C^{1}}+|v|_{C^{1}}$, and the norm
$|U|_{X}=|u|_{\infty}+|v|_{\infty}$ in $C$, where $U=(u,v)$. Here конус $P$ is given
by
$$
P=\{(u,v)\in X:\;u\ge 0, v\ge 0\;\; {\rm for \;\;all}\;\; x\in\Omega\}.  \eqno(2)
$$
So, if $u\neq 0$, $v\neq 0$ belong to $P$, then $-u, -v$ does not belong.
We will work with classical spaces on the intervals
$\bar{I}=[a,b]$, $\hat{I}=]a,b],$ $I=(a,b)$:\\
$C(\bar{I})$ with norm $\parallel u\parallel_{\infty}=max\{|u(x)|:\;x\in\bar{I}\}$;\\
$C^{1}(\bar{I})=\parallel u\parallel_{\infty}+\parallel u'\parallel_{\infty}$;\\
$C_{loc}(I)$, which contains all functions that are locally absolutely
continuous in $I$. We introduce a space $C_{loc}(I)$ because the problem (I)
is singular for $\varphi=0$. The order $\le$ in cone $P$ is understood in the
weak sense, i.e. $y$ is increasing if $a\le b$ implies $y(a)\le y(b)$ and
$y$ is decreasing if $a\le b$ implies $y(a)\ge y(b)$.

{\bf Theorem 1.}[5] (comparison principle in cone) {\it Let $y\in C(\bar{I})
\bigcap C_{loc}(I)$. The function $f$ is defined on $I\times R$. Let $f(x,y)$
is increasing in $y$ function, then
$$
v''-f(x,v)\ge w''-f(x,w)\;\;   a.e.\;\; on\;\;I,  \eqno(3)
$$
$$
v(a)\le w(a), \;\;\;v(b)\le w(b)
$$
implies}
$$
v\le w \;\;\; on\;\;\bar{I}.
$$
Remark 1. Let $f(x,y)$ is decreasing, then theorem 1 remains without changes, if
both parts of (3) multiply onto -1.

For the convenience of defining an ordering relation in cone $P$ we make a
transformation for the problem (I). Let $F(\varphi,a)$ and $G(\varphi,a)$
be defined by (I). Then throuth the transformation $\varphi=-u$ the problem
(I) is reduced to the form
$$
-\frac{d^{2}u}{dx^{2}}=j_{x}\frac{1-u}{\sqrt{(1-u)^{2}-1-a^{2}}}
\stackrel{\triangle}{=}\tilde{F}(j_{x},u,a),\;\;u(0)=0, \;\;u(1)=\varphi_{L},
$$
$$
\eqno(II)
$$
$$
\frac{d^{2}a}{dx^{2}}=j_{x}\frac{a}{\sqrt{(1-u)^{2}-1-a^{2}}}\stackrel{\triangle}{=}
\tilde{G}(j_{x},u,a), \;\;a(0)=0, \;\;a(1)=a_{L}.
$$
We note that all solutions of the problem (I), as well the problem (II), are
symmetric with respect to the transformation of sign for the magnetic
potential $a: (\varphi,a)=(\varphi,-a)$ or the same $(u,a)=(u,-a)$.
Thus we must search only positive solutions $\varphi>0$, $a>0$ in cone $P$
or only negative ones: $\varphi<0$, $a<0$. Thanks to the symmetry of problem
it is equivalently and does not yields the extension of the types of sign-
defined solutions of the problem (I) (respect.(II)). Once more, we note that
introduction of negative electrostatic potential in problem (II) is connected
with more convenient relation between order in cone and positiveness of Green
function for operator $-u''$ that we use below.

{\bf Definition 2.} A pair $[(\varphi_{0},a_{0}), (\varphi^{0},a^{0})]$
is called\\
a) sub-super solution of the problem (I) relative to $P$, if the following
conditions are satisfied
$$
\left\{\begin{array}{ll}
(\varphi_{0},a_{0})\in C_{loc}(I)\bigcap C(\bar{I})\times C_{loc}(I)
\bigcap C(\bar{I}), \\ [0.2cm]
(\varphi^{0},a^{0})\in C_{loc}(I)\bigcap C(\bar{I})\times C_{loc}(I)\bigcap C(\bar{I})
\end{array}\right.; \eqno(4)
$$
$$
\varphi_{0}^{''}-j_{x}\frac{1+\varphi_{0}}{\sqrt{(1+\varphi_{0})^{2}-1-a^{2}}}\stackrel{\triangle}{=}
F(\varphi_{0},a)\le 0\;\;{\rm in}\;\;I, \hspace{3.5cm}
$$
$$
\eqno(5)
$$
$$
(\varphi^{0})''-j_{x}\frac{1+\varphi^{0}}{\sqrt{(1+\varphi^{0})^{2}-1-a^{2}}}\stackrel{\triangle}{=}
F(\varphi^{0},a)\ge 0\;\;{\rm in}\;\;I\;\;\forall a\in[a_{0},a^{0}];
$$
$$
a_{0}^{''}-j_{x}\frac{a_{0}}{\sqrt{(1+\varphi)^{2}-1-a_{0}^{2}}}\stackrel{\triangle}{=}
G(\varphi,a_{0})\le 0\;\;{\rm in}\;\;I, \hspace{3.5cm}
$$
$$
\eqno(6)
$$
$$
(a^{0})''-j_{x}\frac{a^{0}}{\sqrt{(1+\varphi)^{2}-1-(a^{0})^{2}}}\stackrel{\triangle}
{=}G(\varphi,a^{0})\ge 0\;\;{\rm in}\;\;I\;\;\forall \varphi\in[\varphi_{0},\varphi^{0}];
$$
$$
\varphi_{0}\le \varphi^{0}, \;\;\;a_{0}\le a^{0}\;\;{\rm in}\;\;I   \eqno(7)
$$
and on the boundary
$$
\varphi_{0}(0)\le 0\le \varphi^{0}(0),\;\;\;\varphi_{0}(1)\le \varphi_{L}\le \varphi^{0}(1),
$$
$$
\eqno(8)
$$
$$
a_{0}(0)\le 0\le a^{0}(0),\;\;\;a_{0}(1)\le a_{L}\le a^{0}(1);
$$
b) sub-sub solution of the problem (I) relative to $P$, if a condition (4)
is satisfied and
$$
\varphi_{0}^{''}-F(j_{x},\varphi_{0},a_{0})\le 0\;\;{\rm in}\;\;I,
$$
$$
\eqno(9)
$$
$$
a_{0}^{''}-G(j_{x},\varphi_{0},a_{0})\le 0\;\;{\rm in}\;\;I
$$
and on the boundary
$$
\varphi_{0}(0)\le 0, \;\;\varphi_{0}(1)\le\varphi_{L}, \;\;\;a_{0}(0)\le 0, \;\;a_{0}(1)\le a_{L}.
\eqno(10)
$$
Remark 2. In definition 2 the expressions with square roots we take by modulus
$|(1+\varphi)^{2}-1-a^{2}|$.

By analogy with (9), (10) we may introduce the definition of super-super
solution in cone.

{\bf Definition 3.} The functions $\Phi(x,x_{a_{i}},j_{x})$,
$\Phi_{1}(x,x_{\varphi_{j}},j_{x})$ we shall call a {\it semitrivial} solutions
of the problem (I), if $\Phi(x,x_{a_{i}},j_{x})$ is a solution of the scalar
boundary value problem
$$
\varphi''=F(j_{x},\varphi,x_{a_{i}})=j_{x}\frac{1+\varphi}{\sqrt{(1+\varphi)^{2}-1-(x_{a_{i}})^{2}}},
\eqno(III)
$$
$$
\varphi(0)=0, \;\;\varphi(1)=\varphi_{L},
$$
and $\Phi_{1}(x,x_{\varphi_{j}},j_{x})$ is a solution of the scalar boundary
value problem
$$
a''=G(j_{x},x_{\varphi_{j}},a)=j_{x}\frac{a}{\sqrt{(1+x_{\varphi_{j}})^{2}-1-a^{2}}},
\eqno(IV)
$$
$$
a(0)=0, \;\;a(1)=a_{L}.
$$
Here $x_{a_{i}}$, $i=1,2,3$ and $x_{\varphi_{j}}$, $j=1,2$ are respectively, the
{\it indicators} of semitrivial solutions $\Phi(x,x_{a_{i}},j_{x})$,
$\Phi_{1}(x,x_{\varphi_{j}},j_{x})$ defined by the following way:\\
$x_{a_{1}}=0$, if $a(x)=0$;\\
$x_{a_{2}}=a^{0}$, if $a=a^{0}$ be upper solution of the problem (IV);\\
$x_{a_{3}}=a_{0}$, if $a=a_{0}$  be lower solution of the problem (IV);\\
$x_{\varphi_{1}}=\varphi^{0}$, if $\varphi=\varphi^{0}$ be upper solution
of the problem (III);\\
$x_{\varphi_{2}}=\varphi_{0}$, if $\varphi=\varphi_{0}$  be lower solution
of the problem (III).

From definition 3 we obtain the following types of scalar boundary value
problems for semitrivial (in sense of definition 3) solutions of (I)
(resp.(II)):
$$
\varphi''=F(\varphi,0)=j_{x}\frac{1+\varphi}{\sqrt{(1+\varphi)^{2}-1}}, \;\;
\varphi(0)=0,\;\;\varphi(1)=\varphi_{L}.
\hspace{2cm} \eqno({\rm A}_{1})
$$
$$
\varphi''=F(\varphi,a^{0})=j_{x}\frac{1+\varphi}{\sqrt{(1+\varphi)^{2}-1-(a^{0})^{2}}}, \;\;
\varphi(0)=0,\;\;\varphi(1)=\varphi_{L}.  \eqno({\rm A}_{2})
$$
$$
\varphi''=F(\varphi,a_{0})=j_{x}\frac{1+\varphi}{\sqrt{(1+\varphi)^{2}-1-(a_{0})^{2}}}, \;\;
\varphi(0)=0,\;\;\varphi(1)=\varphi_{L}.  \eqno({\rm A}_{3})
$$
$$
a''=G(\varphi^{0},a)=j_{x}\frac{a}{\sqrt{(1+\varphi^{0})^{2}-1-a^{2}}}, \;\;a(0)=0, \;\;
a(1)=a_{L}.  \eqno({\rm A}_{4})
$$
$$
a''=G(\varphi_{0},a)=j_{x}\frac{a}{\sqrt{(1+\varphi_{0})^{2}-1-a^{2}}}, \;\;a(0)=0, \;\;a(1)=a_{L}.
\eqno({\rm A}_{5})
$$

We shall find the solutions of problems $(A_{1})-(A_{3})$ with condition
$$
\varphi_{0}<\varphi^{0},
$$
where $\varphi_{0}(x_{a_{1}})$, $\varphi^{0}(x_{a_{2}})$ are respectively, lower and upper
solutions of problem $(A_{1})$. The solution $(\varphi,a)$ of problem (I)
should be belong to the interval
$$
\varphi\in\Phi(\varphi,0)\bigcap\Phi(\varphi,a^{0})\bigcap\Phi(\varphi,a_{0}),
$$
$$
a\in\Phi_{1}(\varphi^{0},a)\bigcap\Phi_{1}(\varphi_{0},a).
$$
Moreover, the ordering of lower and upper solutions of problems $(A_{1})-
(A_{3})$ is satisfied
$$
\varphi_{0}(x_{a_{1}})<\varphi_{0}(x_{a_{2}})<\varphi_{0}(x_{a_{3}})<
\varphi^{0}(x_{a_{2}})<\varphi^{0}(x_{a_{1}}).
$$
We shall seek the solution of problems $(A_{4})-(A_{5})$ with condition
$$
a_{0}<a^{0},
$$
in this case the following ordering of lower and upper solutions of problems
$(A_{4})-(A_{5})$
$$
a_{0}(x_{\varphi_{1}})<a_{0}(x_{\varphi_{2}})<a^{0}(x_{\varphi_{2}})<a^{0}(x_{\varphi_{1}}).
$$
is satisfied.

We go over to the direct study of the problem (III), which includes a cases
$(A_{1})-(A_{3})$. Let us consider the boundary value problem (III) with
$$
F(x,\varphi):(0,1]\times(0,\infty)\rightarrow (0,\infty).  \eqno({\rm B}_{1})
$$
In condition $(B_{1})$ for $F(x,\varphi)$ we dropped index $a_{i}$, considering
a general case of nonlinear dependence $F$ of $x$.

We shall assume that $F$ is Caratheodory function, i.e.
$$
F(\cdot,s)\;\;\;{\rm is\;\;measurable\;\;for\;\;all}\;\;s\in R,    \eqno({\rm B}_{2})
$$
$$
F(x,\cdot) \;\;\;{\rm is\;\;continuous\;\; a.e.}\;\;{\rm for}\;\;x\in]0,1], \eqno({\rm B}_{3})
$$
and the following conditions hold
$$
\int_{0}^{1}s(1-s)Fds<\infty. \eqno({\rm B}_{4})
$$
$$
\partial F/\partial \varphi>0, \;\;{\rm i.e.}\;\;F\;{\rm is\;\;increasing\;\;in}\;
\varphi.
\eqno({\rm B}_{5})
$$
There are $\gamma(x)\in L^{1}(]0,1])$ and $\alpha\in R$, $0<\alpha<1$ such that
$$
|F(x,s)|\le\gamma(x)(1+|s|^{-\alpha}), \;\;\forall (x,s)\in ]0,1]\times R.
\eqno({\rm B}_{6})
$$

We are interested in a positive classical solution of equation (III), i.e.
$\varphi>0$ in $P$ for $x\in]0,1]$ and $\varphi\in C([0,1])\bigcap C^{2}(]0,1])$.
The problem (III) is singular, therefore, condition $(B_{1})$ is not
fulfilled on the interval $\varphi\in(0,\infty)$ and in this connection,
the well-known theorems [3] on existence of lower and upper solution in cone
$P$ does not work. It follows from theorem 1, since $F$ in (III) is
increasing in $\varphi$, then $\varphi<w$ for $x\in]0,1]$, where $\varphi$
and $w$ satisfy to the diffeential inequality (3).

{\bf Theorem 2.} {\it Assume conditions $(B_{2})-(B_{6})$. Then there exists
a positive solution $\varphi\in C([0,1])\bigcap C^{2}(]0,1])$ to the boundary
value problem}(III).

Proof. Let  $\varphi>0$ is a solution of problem (III).
By theorem 1 $\varphi<w$ for $x\in]0,1]$. Take $\epsilon>0$ and consider
equation
$$
\varphi_{\epsilon}^{''}=j_{x}\frac{1+\varphi_{\epsilon}+\epsilon}{\sqrt{
(1+\varphi_{\epsilon}+\epsilon)^{2}-1-(x_{a_{i}})^{2}}}\stackrel{\triangle}
{=}F_{\epsilon}(j_{x},\varphi_{\epsilon}+\epsilon,x_{a_{i}}).
$$
$$
\eqno(11)
$$
$$
\varphi_{\epsilon}(0)=0, \;\;\;\varphi_{\epsilon}(1)=\varphi_{L}.
$$
Let $w$ and $\varphi$ are upper and lower solutions of equation (11)
(below, in proposition 1 is shown that such solutions really exist).
Hence the theorem on monotone iterations [6] gives an existence of classical
solution $\varphi_{\epsilon}$ of equation (11), which satisfies $w>\varphi_{
\epsilon}>\varphi$ for $x\in]0,1]$ and is bounded in $C$. Thus
$F_{\epsilon}(j_{x},\varphi_{\epsilon}+\epsilon,x_{a_{i}})$ is bounded
and there exists uniform limit $\lim_{\epsilon\rightarrow 0}\varphi_{\epsilon}=
\varphi$. It follows from the last, if $0<\eta<\frac{1}{2}$, then
$\lim_{\epsilon\rightarrow 0}F_{\epsilon}(j_{x},\varphi_{\epsilon}+\epsilon,x_{a_{i}})=
F(j_{x},\varphi,x_{a_{i}})$ uniformly on $[\eta, 1-\eta]$ and $\varphi>0$ for
$x\in[\eta,1-\eta]$.

Since $\varphi_{\epsilon}$ converges uniformly on $[0,1]$, this is implies
the existence $\lim_{\epsilon\rightarrow 0}\varphi_{\epsilon}^{'}(\eta)$.
Therefore there exists $\lim_{\epsilon\rightarrow 0}\varphi_{\epsilon}^{''}(x)$
on compact subsets (0,1) and $\{\varphi_{\epsilon}^{'}\}$
uniformly converges on (0,1) to the differentiable function $\varphi^{'}$ on
$[\eta,1-\eta]$. It follows from the last, $\varphi$ is twice differentiable on
$[\eta,1-\eta]$, $\varphi^{''}=F(j_{x},\varphi,x_{a_{i}})$, $x\in[\eta,1-\eta]$
and $u\in C([0,1])\bigcap C^{2}(]0,1])$ is a positive solution of problem (III).

Remark 3. Delicate moment in proof of theorem 2 is connected with finding of a
lower $\varphi$ and an upper $w$ solutions for perturbated problem (11).
As a lower solution we can take solution of equation $(A_{1})$
(semitrivial solution $\varphi$), then an upper solution will be, for example,
maximal solution of equation $(A_{1})$.

Application of monotone iteration techniques to the equation (III) gives an
existence of maximal solution $\bar{\varphi}(x,j_{x})$ such that
$$
\varphi(x,x_{j})\le\bar{\varphi}(x,x_{j})<w(x)\;\;{\rm для}\;\;x\in]0,1].
\eqno(12)
$$

{\bf Proposition 1.} {\it Let $0<c\le j_{x}\le j_{x}^{max}$. Then equation
$(A_{1})$
$$
\varphi^{''}=F(j_{x},\varphi,0)=j_{x}\frac{1+\varphi}{\sqrt{\varphi(2+\varphi)}},
$$
$$
\varphi(0)=0, \;\;\varphi(1)=\varphi_{L}
$$
has a lower positive solution
$$
u_{0}=\delta^{2}x^{4/3}, \eqno(13)
$$
if
$$
4\delta^{3}\ge 9j_{x}^{max}(1+\delta^{2})/\sqrt{2+\delta^{2}}  \eqno(14)
$$
and an upper positive solution
$$
u^{0}=\alpha+\beta x\;\;(\alpha,\beta>0),  \eqno(15)
$$
here
$$
\varphi_{L}\ge\delta^{2},   \eqno(16)
$$
where $\delta$ is defined from} (14).

Remark 4. Square root is taking as $\sqrt{|\varphi(2+\varphi)|}$ in the case of
negative solutions. Here $u^{0}=-\epsilon x$ is an upper solution, and
$u_{0}=-2+\epsilon$ is a lower solution $(0<\epsilon<1)$. Hence,
equation $(A_{1})$ has negative solution only for $0<\varphi_{L}<-2$ because
$F(x,-2)=-\infty$.

It follows from (14), (16) that a value of current is limited by the value of
electrostatic potential on the anode $\varphi_{L}$
$$
j_{x}\le j_{x}^{max}\le {\cal F}(\varphi_{L}). \eqno(17)
$$
Analysis of lower and upper solutions (13), (15) exhibits that for $\delta^{2}=
\varphi_{L}>2$ and $\alpha=\beta\le 1$ interval in $x$ between lower and upper
solutions is decreased, and  for the large values of potential $\varphi_{L}$
diode makes on regime $\varphi_{L}x^{4/3}$.

{\bf Proposition 2.} {\it Let $0<c\le j_{x}\le j_{x}^{max}$. Then equation
$(A_{4})$
$$
a^{''}=G(j_{x},\varphi^{0},a)=j_{x}\frac{a}{\sqrt{(1+\varphi^{0})^{2}-1-a^{2}}}, \;\;
a(0)=0,\;\;a(1)=a_{L}
$$
with a lower solution $a_{0}=0$ and an upper solution $a^{0}=u^{0}>0$,
conditions (14), (16) has an unique solution $a(x,j_{x},c)$, which is positive,
moreover}
$$
0\le a_{L}\le\sqrt{\varphi^{0}(2+\varphi)}.  \eqno(18)
$$
Proof. The positive solution of problem $(A_{4})$ is concave and be found as
a solution of initial problem with $a(0)=0, \;\;a'(0)=c$, where $c$ is shooting
parameter. The solution $a=a(x,j_{x},c)$ is unique and strongly decreasing in
$c$ because right part of differential equation is decreasing in $a$.
The least nonnegative solution is $f(x,j_{x},0)=0$ and for
$0\le a_{L}\le\sqrt{\varphi^{0}_{L}(2+\varphi^{0}_{L})}$ exists only one
solution and no positive solutions for other values $a_{L}$.

Remark 5. The problem $(A_{5})$ is considered by analogy with problem $(A_{4})$,
change of an upper solution $a^{0}=u^{0}$ to a lower  $a^{0}=u_{0}$ one and
$0\le a_{L}\le\sqrt{\varphi_{0L}(2+\varphi_{0L})}$.

Following to the definition 2 and propositions 1, 2, solutions of problems
(III), (IV) we can write in the form (fig.1):

{\small
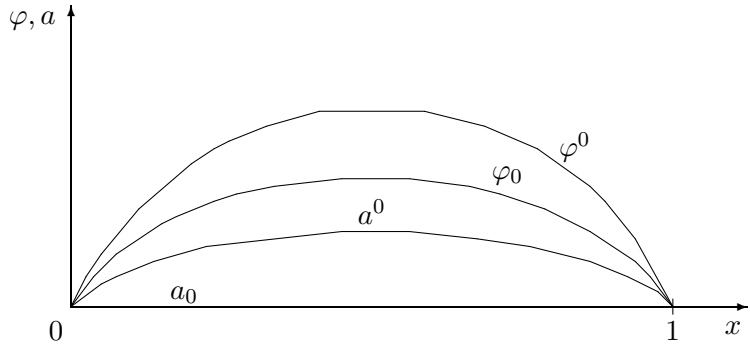
\begin{figure}[h]
\unitlength=1.00mm
\special{em:linewidth 0.4pt}
\linethickness{0.4pt}
\begin{picture}(159.00,64.00)
\put(40.00,5.00){\vector(1,0){90.00}}
\put(40.00,5.00){\vector(0,1){40.00}}
\put(38.00,3.00){\makebox(0,0)[ct]{$0$}}
\put(128.00,3.00){\makebox(0,0)[ct]{$x$}}
\put(38.00,43.00){\makebox(0,0)[rc]{$\varphi, a$}}
\emline{120.00}{6.00}{1}{120.00}{4.00}{2}
\put(120.00,3.00){\makebox(0,0)[ct]{$1$}}
\emline{40.00}{5.00}{3}{43.00}{9.00}{4}
\emline{43.00}{9.00}{5}{46.00}{12.00}{6}
\emline{46.00}{12.00}{7}{52.00}{16.00}{8}
\emline{52.00}{16.00}{9}{54.00}{17.00}{10}
\emline{54.00}{17.00}{11}{59.00}{19.00}{12}
\emline{59.00}{19.00}{13}{62.00}{20.00}{14}
\emline{62.00}{20.00}{15}{67.00}{21.00}{16}
\emline{67.00}{21.00}{17}{76.00}{22.00}{18}
\emline{76.00}{22.00}{19}{85.00}{22.00}{20}
\emline{85.00}{22.00}{21}{93.00}{21.00}{22}
\emline{93.00}{21.00}{23}{97.00}{20.00}{24}
\emline{97.00}{20.00}{25}{103.00}{18.00}{26}
\emline{103.00}{18.00}{27}{109.00}{15.00}{28}
\emline{109.00}{15.00}{29}{115.00}{11.00}{30}
\emline{115.00}{11.00}{31}{117.00}{9.00}{32}
\emline{117.00}{9.00}{33}{120.00}{5.00}{34}
\emline{40.00}{5.00}{35}{42.00}{9.00}{36}
\emline{42.00}{9.00}{37}{44.00}{12.00}{38}
\emline{44.00}{12.00}{39}{49.00}{18.00}{40}
\emline{49.00}{18.00}{41}{56.00}{24.00}{42}
\emline{56.00}{24.00}{43}{59.00}{26.00}{44}
\emline{59.00}{26.00}{45}{61.00}{27.00}{46}
\emline{61.00}{27.00}{47}{66.00}{29.00}{48}
\emline{66.00}{29.00}{49}{73.00}{31.00}{50}
\emline{159.00}{64.00}{51}{159.00}{64.00}{52}
\emline{159.00}{64.00}{53}{159.00}{64.00}{54}
\emline{40.00}{5.00}{55}{44.00}{8.00}{56}
\emline{44.00}{8.00}{57}{46.00}{9.00}{58}
\emline{46.00}{9.00}{59}{51.00}{11.00}{60}
\emline{51.00}{11.00}{61}{58.00}{13.00}{62}
\emline{58.00}{13.00}{63}{76.00}{15.00}{64}
\emline{76.00}{15.00}{65}{85.00}{15.00}{66}
\emline{85.00}{15.00}{67}{94.00}{14.00}{68}
\emline{94.00}{14.00}{69}{101.00}{13.00}{70}
\emline{101.00}{13.00}{71}{109.00}{11.00}{72}
\emline{109.00}{11.00}{73}{114.00}{9.00}{74}
\emline{114.00}{9.00}{75}{118.00}{7.00}{76}
\emline{118.00}{7.00}{77}{120.00}{5.00}{78}
\emline{73.00}{31.00}{79}{87.00}{31.00}{80}
\emline{87.00}{31.00}{81}{95.00}{29.00}{82}
\emline{95.00}{29.00}{83}{102.00}{26.00}{84}
\emline{102.00}{26.00}{85}{109.00}{21.00}{86}
\emline{109.00}{21.00}{87}{111.00}{19.00}{88}
\emline{111.00}{19.00}{89}{115.00}{14.00}{90}
\emline{115.00}{14.00}{91}{120.00}{5.00}{92}
\put(98.00,22.00){\makebox(0,0)[cb]{$\varphi_{0}$}}
\put(80.00,16.00){\makebox(0,0)[cb]{$a^0$}}
\put(55.00,6.00){\makebox(0,0)[cb]{$a_0$}}
\put(107.00,25.00){\makebox(0,0)[cb]{$\varphi^0$}}
\end{picture}
\caption{location of lower $(\varphi_{0},a_{0})$ and upper $(\varphi^{0},a^{0})$
solutions}
\end{figure}
}

\noindent
lower-lower $(\varphi_{0},a_{0})$):
$$
\varphi_{0}=u_{0}=\delta^{2}x^{4/3}, \;\;a_{0}=0, \;\;\varphi_{L}\ge\delta^{2};
$$
upper-lower $(\varphi^{0}, a_{0})$:
$$
\varphi^{0}=u^{0}=\alpha+\beta x, \;\;a_{0}=0, \;\;\delta^{2}\le\varphi_{L}
\le {\cal C}, \;\;{\cal C}=max\{\alpha,\beta\};
$$
lower-upper $(\varphi_{0}, a^{0}$):
$$
\varphi_{0}=u_{0}=\delta^{2}x^{4/3}, \;\;a^{0}=u^{0}, \;\;\varphi_{L}\ge \delta^{2}, \;\;
a_{L}\le \sqrt{(u_{0}(2+u_{0})};
$$
upper-upper  $(\varphi^{0}, a^{0})$:
$$
\varphi^{0}=u^{0}=\alpha+\beta x, \;\;a^{0}=u^{0}, \;\;\varphi_{L}\le {\cal C},
a_{L}\le a^{0}\le u^{0}.
$$

{\bf 3. Existence of solutions of system (I).}

In the previous section we demonstrated the existence of semitrivial solutions
of system (I). Here we show existence of solutions for the complete system (I),
using the following McKenna-Walter theorem.

{\bf Theorem 3.}[4] {\it Assume conditions $(B_{1})-(B_{6})$. We assume that
there exists the ordered pair $(\underline{u},\bar{u})$ $-$ lower and upper
solutions, i.e.
$$
\underline{u}, \bar{u}\in C_{loc}((0,1])^{2}\bigcap C([0,1])^{2}, \;\;\underline{u}
\le\bar{u}\;\;{\rm в}\;\;]0,1]
$$
$$
\underline{u}(0)\le 0\le\bar{u}(0), \;\;\underline{u}(1)\le u_{L}\le\bar{u}(1); \;\;
u_{L}\stackrel{\triangle}{=}(\varphi_{L},a_{L}),
$$
$$
\forall x\in ]0,1]: \;\forall z\in R^{2},
$$
$$
\underline{u}(x)\le z\le\bar{u}(x), \;\;z_{k}=\underline{u}_{k}(x);
$$
$$
-\underline{u}^{''}_{k}(x)\ge h_{k}(x,z) \eqno(19)
$$
and
$$
\forall x\in ]0,1]:\; \forall z\in R^{2},
$$
$$
\underline{u}(x)\le z\le\bar{u}(x), \;\;z_{k}=\bar{u}_{k}(x):
$$
$$
-\bar{u}^{''}_{k}(x)\le h_{k}(x,z)  \eqno(20)
$$
for all $k\in\{1,2\}$. Then there exists a solution $u\in C^{2}((0,1])^{2}\bigcap
C([0,1])^{2}$ of the problem}
$$
-u''=h(\cdot, u(\cdot))\;\;{\rm на}\;\;]0,1]
$$
$$
u(0)=0, \;\;u(1)=u_{L}.
$$

For keeping of ordering of lower and upper solutions in theorem 3
(in cone $P$) we write differential inequalities (19), (20) in the
following form
$$
\forall z\in [v(x),w(x)],\;\;z_{1}=w_{1}(x):
$$
$$
^{\pm}w_{1}^{''}(x)\stackrel{(\ge)}{\le}\;^{\pm}F_{1}(w_{1}(x),z_{2})
$$
$$
\forall z\in[v(x),w(x)], \;\;z_{1}=v_{1}(x):
$$
$$
^{\pm}v_{1}^{''}(x)\stackrel{(\le)}{\ge}\;^{\pm}F_{1}(v_{1}(x),z_{2})
$$
$$
\forall z\in[v(x),w(x)]; \;\;z_{2}=w_{2}(x)
$$
$$
^{\pm}w_{2}^{''}(x)\stackrel{(\ge)}{\le}\;^{\pm}F_{2}(z_{1},w_{2})
$$
$$
\forall z\in[v(x),w(x)]; \;\;z_{2}=v_{2}(x)
$$
$$
^{\pm}v_{2}^{''}(x)\stackrel{(\le)}{\ge}\;^{\pm}F_{2}(z_{1},v_{2}).
$$

Remark 6. Change of signs with (+) to ($-$) in differential inequalities is
connected with adjustment of signs and ordering $(\le)$ of lower (upper)
solutions of system (I) in definition 2 and lower (upper) solutions in
theorem 3.

From the last relations we obtain
$$
\left\{ \begin{array}{ll}
w^{''}(x)=F_{1}(w_{1}(x),0)\le F_{1}(w_{1},z_{2}) \\ [0.2cm]
v_{1}^{''}(x)\ge\sup\limits_{z_{2}}F_{1}(v_{1}(x),z_{2})
\end{array}\right.,
\eqno(21)
$$
$$
\left\{ \begin{array}{ll}
w_{2}^{''}(x)\le F_{2}(z_{1},w_{2})\\[0.2cm]
v_{2}^{''}(x)\ge \sup\limits_{z_{1}} F_{2}(z_{1},v_{2})
\end{array}\right.. \hspace{2.5cm}
\eqno(22)
$$
From inequality $v_{2}^{''}(x)\ge\sup_{z_{1}}F_{2}(z_{1},v_{2})$ we get
estimations to the value of magnetic field on the anode $a_{L}$
$$
a_{L}\le\frac{j_{x}}{2}\le\frac{j_{x}^{max}}{2}\le\frac{{\cal F}(\varphi_{L})}
{2} \eqno(23)
$$
taking account of (17) and $\Theta_{L}>0$. Under realization of estimation (23)
diode works in "noninsulated" regime, moreover the value $a_{L}$ is limited by
value of electrostatic potential on the anode $\varphi_{L}$ with critical value
$\varphi_{L}=2$. In increasing of magnetic potential $a_{L}$ diode transfers in
"isolated" regime that leads to more complicated problem with free boundary.

Thus we have the following main result of this paper.

{\bf Theorem 4.} {\it Assume conditions $(B_{2})$, $(B_{3})$, $(B_{6})$ and
inequalities (14), (17), (23). Then the problem (I) possesses a positive
solution  $P$ such that
$$
\left\{\begin{array}{ll}
\varphi_{0}^{''}\ge j_{x}F(\varphi_{0},z_{2}), \;\;z_{2}\in[0,\varphi^{0}]\\ [0.2cm]
(\varphi^{0})''\le j_{x}F(\varphi^{0},z_{2}), \;\;z_{2}\in[0,\varphi^{0}]
\end{array}\right.,
$$
$$
\left\{\begin{array}{ll}
a_{0}^{''}\ge G(j_{x},z_{1},a_{0}), \;\;z_{1}\in[\varphi_{0},\varphi^{0}]\\[0.2cm]
(a^{0})''\le G(j_{x},z_{1},a^{0}), \;\;z_{1}\in[\varphi_{0},\varphi^{0}]
\end{array}\right.,
$$
where $\varphi_{0}=\delta^{2} x^{4/3}$ is a lower solution of problem
$(A_{1})$, $\varphi^{0}=\alpha+\beta x\;(\alpha,\beta>0)$ is an upper
solution of problem $(A_{1})$ with condition
$\varphi_{L}\ge\delta^{2}$; $a_{0}=0$ is a lower solution of problem $(A_{4})$
with condition} $0\le a_{L}\le\sqrt{\varphi^{0}(2+\varphi^{0})}$.

{\bf Acknowledgments}. The author would like to thank J.Batt, P.Degond and
N.Ben Abdallah for collaboration.

\end{document}